\documentclass[aps,pra,twocolumn,10pt,superscriptaddress]{revtex4-1}

\usepackage{amsmath, amssymb}
\usepackage{mathtools}
\usepackage{braket}
\usepackage{siunitx}

\usepackage{math}
\usepackage{color}

\usepackage{hyperref}

\begin{document}

\title{The effect of bound state dressing in laser assisted radiative recombination}

\author{Robert A. M\"uller}
\affiliation{Helmholtz-Institute Jena, 07743 Jena, Germany}
\affiliation{Friedrich-Schiller-Universit\"at Jena, 07743 Jena, Germany}
\author{Daniel Seipt}
\affiliation{Helmholtz-Institute Jena, 07743 Jena, Germany}
\author{Stephan Fritzsche}
\affiliation{Helmholtz-Institute Jena, 07743 Jena, Germany}
\affiliation{Friedrich-Schiller-Universit\"at Jena, 07743 Jena, Germany}
\author{Andrey Surzhykov}
\affiliation{Helmholtz-Institute Jena, 07743 Jena, Germany}

\date{\today}

\begin{abstract}
We present a theoretical study on the recombination of a free electron into the ground state of a hydrogen-like ion in the presence of an external laser field. Emphasis is placed on the effects caused by the laser dressing of the residual ionic bound state. To investigate how this dressing affects the total and angle-differential cross section of laser assisted radiative recombination (LARR) we apply first-order perturbation theory and the separable Coulomb-Volkov-continuum ansatz. Using this approach detailed calculations were performed for low-$Z$ hydrogen like ions and laser intensities in the range from $I_L=\SI{e12}{W/cm^2}$ to $I_L=\SI{e13}{W/cm^2}$. It is seen that the total cross section as a function of the laser intensity is remarkably affected by the bound state dressing. Moreover the laser dressing becomes manifest as asymmetries in the angular distribution and the (energy) spectrum of the emitted recombination photons.
\end{abstract}

\maketitle

\section{Introduction}
\label{sec:introduction}
The recombination of a continuum electron into a bound state of an atom or ion accompanied by the emission of a photon is commonly called \emph{radiative recombination} (RR). This fundamental process can be observed in many laboratory and astrophysical plasmas and in ion or electron storage rings \cite{yamaguchi_discovery_2009,schuch_radiative_1993, hahn_electron-ion_1997}. The cross section of this process as well as the energy and angular distribution of the emitted recombination photons have been studied extensively during the last decades both in theory and experiment (see Ref. \cite{surzhykov_photonphoton_2002,fritzsche_radiative_2005,hahn_electron-ion_1997} and references therein). The characteristics of the emitted radiation may change remarkably after all if the electron and the nucleus are exposed to an external laser field. This additional field can accelerate or decelerate the electron before the recombination takes place and, hence, may broaden the spectrum and also modify the emission pattern of the recombination photons. It is of major importance to understand these effects since \emph{laser assisted radiative recombination} (LARR) is the third step in the process of high harmonic generation \cite{corkum_plasma_1993,lewenstein_theory_1994}. Moreover the external laser can cause a significant gain of the recombination yield which may support the generation of neutral antimatter \cite{rogelstad_stimulated_1997,wesdorp_field-induced_2000,shuman_multiphoton_2008}.

LARR was studied in a number of experiments e.g. at the test storage ring (TSR) in Heidelberg, at the GSI in Darmstadt and at the Helmholtz-Institute Jena \cite{schramm_observation_1991, scrinzi_laser_1995, schramm_laser-induced_1996,mohamed_effects_2011,moller_dependence_2012}. To understand the evidences found in these experiments a number of theoretical studies have been performed \cite{keldysh_ionization_1965, faisal_multiple_1973, reiss_effect_1980, corkum_plasma_1993, lewenstein_theory_1994}. Up to now most of these studies were carried out in the context of the \emph{strong field approximation} (SFA). In this theory, originally introduced by Keldysch, Faisal and Reiss \cite{keldysh_ionization_1965, faisal_multiple_1973, reiss_effect_1980}, the incident electron moves solely in the field of the external laser while the laser influence is neglected for the final bound electron. During the recent decades some advancements of the SFA approach were proposed. In particular an extension was developed to include the influence of the Coulomb field into the description of the incident electron \cite{jaron_stimulated_2000}. More recently first steps to include the laser influence on the residual bound state in the theory either perturbatively or by means of an Floquet expansion have been done \cite{bivona_polarization_2007,li_theory_2009,zheltukhin_effects_2011, zheltukhin_resonant_2012}. Combining these advancements Shchedrin and Volberg \cite{shchedrin_analytical_2011} performed a technical analysis of a theory where both, the Coulomb distorsion of the continuum wave function and the perturbative dressing of the residual bound state were included. However up to now no calculations of the differential LARR cross section have been presented respecting the influence of both potentials on both electron states. Moreover the effect of bound state dressing on two-color photoionization has been discussed \cite{cionga_target_1993} but was not analyzed in the context of LARR up to now. Therefore, in this contribution we investigate in detail, how the laser dressing of the bound state affects the photon emission in LARR. Our study is based on an S-matrix approach where we account for the influence of both fields (laser and Coulomb) in the representation of the initial and final electron state \cite{li_theory_2009,shchedrin_analytical_2011}. The incident electron is described by a separable Coulomb-Volkov wave function, the final one by a perturbative expansion of the laser dressed bound state up to the first order. Using this approach,  outlined in Sec. \ref{sec:theory}, we present results for the angle-differential partial cross section in Sec. \ref{ssec:results differential} as well as the total cross section in Sec. \ref{ssec:results total}. We find that the total cross section is typically enhanced due to the assisting laser field. Moreover the bound state dressing significantly affects the angular as well as the energy distribution of the recombination photons.

Atomic units are used throughout this paper ($e=m_e=\hbar=1$), except when stated otherwise.
\section{Geometry of laser assisted radiative recombination}
\label{sec:geometry}
\begin{figure}[tb]
\centering
\includegraphics[width=\linewidth]{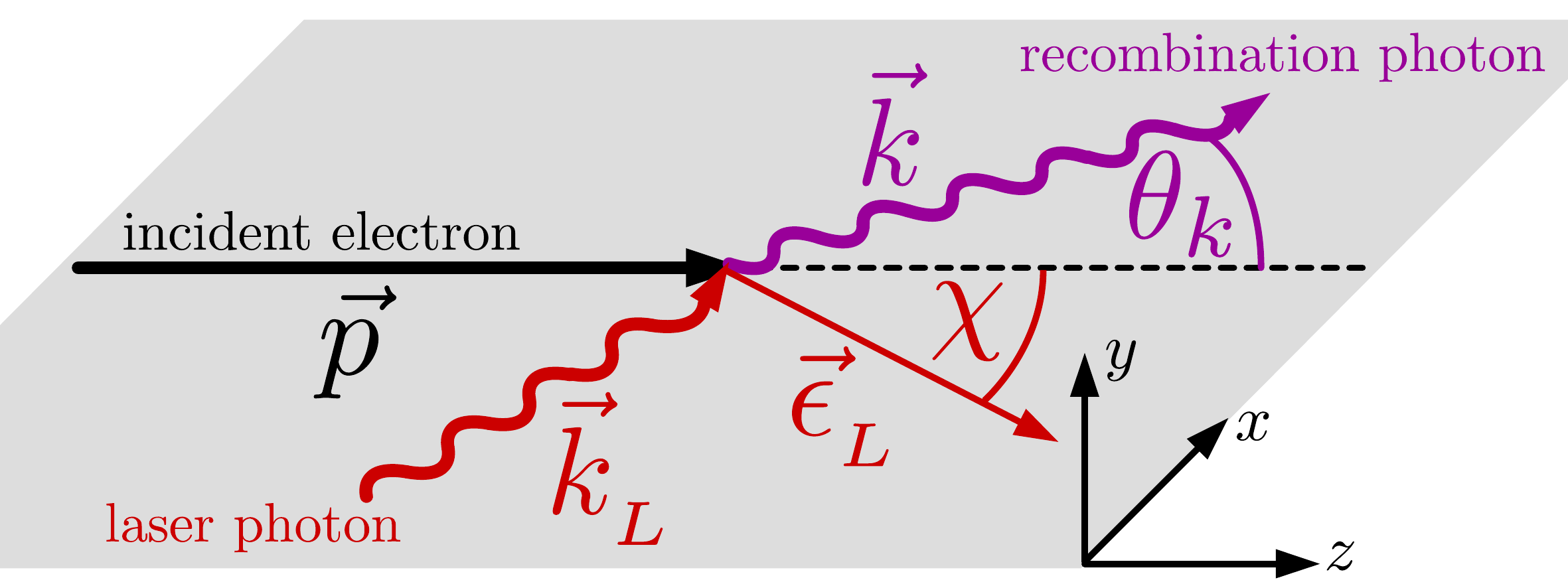}
\caption{Geometry of laser assisted radiative recombination. The $z$-axis is given by the asymptotic momentum $\vp$ of the incoming electron. The $xz$-plane is defined by $\vp$ and the laser polarization $\veps_L$. The angle between these vectors is $\chi$. The recombination photon is assumed to be emitted in the $xz$-plane and therefore characterized by one angle $\theta_k$.\label{fig:geometry}}
\end{figure}

Let us first discuss the geometry used for our description of LARR. Fig. \ref{fig:geometry} displays an atom in its rest frame. The quantization axis ($z$-axis) is chosen along the asymptotic momentum of the incident electron $\vp$. The polarization vector $\veps_L$ of the laser photon together with the electron momentum $\vp$ defines the $xz$-plane. In this work we restrict ourselves to coplanar geometries in which the propagation direction $\vk$ of the recombination photon lies also in the $xz$-plane. Therefore we can characterize both, the polarization of the laser photon $\veps_L$ and the momentum $\vk$ of the recombination photon solely by their polar angles $\chi$ and $\theta_k$ with respect to the $z$-axis.
\section{Theory}
\label{sec:theory}
In this section we shall derive analytical expressions for the cross section LARR. This cross section can be expressed in terms of the S-matrix element that is a measure for the probability that a continuum electron with momentum $\vp$ recombines into the bound state with quantum numbers $(n,l,m)$ accompanied by the emission of a single recombination photon  with frequency $\omega_k$ and polarization $\veps_k^{\lambda}$:
\begin{equation}
S_{nlm}=-i\int_{-\infty}^{\infty}\dd t \langle\psi_{nlm}(\vr,t)\vert\e^{i(\omega_k t-\vk\cdot\vr)}(\veps^{\lambda}_k\cdot\nabla)\vert\psi_{\vp}(\vr,t)\rangle.
\label{eq:sgeneral}
\end{equation}
Here $\lambda$ is the helicity of the recombination photon and $\psi_{\vp}(\vr,t)$ and $\psi_{nlm}(\vr,t)$ are the initial continuum and final bound state wave functions of the recombining electron. In the non relativistic regime these wave functions are taken as solutions of the time dependent Schr\"odinger equation with the Hamiltonian
\begin{equation}
\hat{H}(t)=\frac{\hat{\vp}^2}{2}-\frac{Z}{r}+\vec{A}_L(t)\cdot\hat{\vp}+\frac{1}{2}\vec{A}_L(t)^2,
\label{eq:hamiltonian}
\end{equation}
where $\hat{\vp}$ is the electron momentum operator, $Z$ the nuclear charge and $\vec{A}_L(t)$ the vector potential of the external laser field with frequency $\omega_L$ and polarization $\veps_L$. If the wavelength $\lambda_L$ of this field is much larger then the size of the atom $a_0$ the dipole approximation can be applied and the vector potential written as
\begin{equation}
\vec{A}_L(t)=\veps_L A_L \cos (\omega_L t).
\end{equation}
The corresponding electric field is then
\begin{equation}
\vec{E}_L(t)=\veps_L E_L \sin (\omega_L t).
\end{equation}
The laser intensity $I_L$ in terms of the electric field amplitude $E_L=\omega_L A_L$ is given by $I_L=c E_L^2/4\pi$.
\subsection{Bound and continuum wave functions}
In order to calculate the S-matrix \eqref{eq:sgeneral} of LARR we need to know the bound and continuum state wave functions $\psi_{nlm}(\vr,t)$ and $\psi_{\vp}(\vr,t)$. Since no analytical solutions are known for the Schr\"odinger equation with Hamiltonian $\hat{H}(t)$ (Eq. \eqref{eq:hamiltonian}) we will show below how approximations can be derived for the initial continuum and final bound state of the electron.
\subsubsection{Separable Coulomb-Volkov continuum state}
In Eq. \eqref{eq:sgeneral} $\psi_{\vp}(\vr,t)$ describes the incident electron moving in the superposition of the atomic and the external laser potential. A well established approach to approximate such a wave function, is to construct so called \emph{separable Coulomb-Volkov continuum} states \cite{jain_compton_1978}. We briefly recall here the derivation of this wave function that has been originally introduced by Jain and Tzoar \cite{jain_compton_1978}.

We assume that the electron motion in the continuum is mainly influenced by the external laser field. Therefore the starting point for the construction of $\psi_{\vp}(\vr,t)$ is the well known Volkov wave function
\begin{equation}
\begin{split}
\chi_{\vp}(\vr,t)=\e^{i\vp\cdot\vr}\exp\left[-i\left(E_i t + \frac{A_L}{\omega_L}\veps_L\cdot\vp\sin (\omega_L t)\right.\right.\\
\left.\left.+\frac{A_L^2}{2}\int_{-\infty}^t\cos^2(\omega_L \tau)\dd \tau\right)\right],
\end{split}
\label{eq:volkov}
\end{equation}
which is the exact solution of the Schr\"odinger equation for an electron with kinetic energy $E_i$ moving in a spatially homogeneous laser field switched on adiabatically at $t_0\rightarrow-\infty$. To additionally account for the relatively weak influence of the Coulomb field we replace the plane term $\exp(i\vp\cdot\vr)$ in Eq. \eqref{eq:volkov} by the continuum wave function $\phi_{\vp}(\vr)$ of a hydrogen-like atom \cite{landau_quantum_1991}
\begin{widetext}
\begin{equation}
\begin{aligned}
\psi_{\vp}(\vr,t)&=\phi_{\vp}(\vr)\exp\left[-i\left(E_i t + \frac{A_L}{\omega_L}\veps_L\cdot\vp\sin (\omega_L t)\right)\right]\\
&=(2\pi)^{-\frac{3}{2}}\exp\left[i\left(\vp\cdot\vr-E_i t-\frac{A_L}{\omega_L}\veps_L\cdot\vp\sin (\omega_L t)\right)\right]\exp\left(\frac{\pi Z}{2p}\right) \Gamma(1-i Z/p){}_1\mathrm{F}_1\left[i Z/p,1,i(pr-\vp\cdot\vr)\right],
\end{aligned}
\label{eq:continuum}
\end{equation}
\end{widetext}
but neglect all terms of the order $\sim A_L^2$ and higher. Here ${}_1\mathrm{F}_1(a,b,z)$ and $\Gamma(z)$ are the confluent hypergeometric function and the gamma function, respectively.

This approximation \eqref{eq:continuum} is valid as long as the asymptotic electron momentum $p$ is large compared to the momentum transferred to the electron by the laser field \cite{jain_compton_1978, cavaliere_particle-atom_1980}. As seen from Eq. \eqref{eq:hamiltonian} this momentum transfer is quantified by the laser vector potential $\vec{A}_L(t)$. Therefore the validity condition reads as
\begin{equation}
\frac{A_L}{p}=\frac{A_L}{\sqrt{2E_i}}\ll 1.
\end{equation}
For further discussion of the separable Coulomb-Volkov wave function and its validity we refer the reader to the literature (e.g. Refs. \cite{jain_compton_1978, cavaliere_particle-atom_1980}).
\subsubsection{Laser dressed bound state}
In the last section we derived the wave function $\psi_{\vp}(\vr,t)$ for a continuum state electron in a combined laser and Coulomb field. We obtained $\psi_{\vp}(\vr,t)$ under the assumption that the atomic field is much weaker than the laser one. This is not the case if the electron is in a bound state. The intensity of the coulomb field for the ground state of hydrogen is $I_C=(\alpha^2 e/a_0m_e)^2 c^5 \epsilon_0/2=\SI{3.5e16}{W/cm^2}$ which is much larger than any laser intensity discussed in this paper. To derive the bound state wave function, therefore, we can treat the laser vector potential $\vec{A}_L(t)$ perturbatively. In first order time dependent perturbation theory $\psi_{nlm}(\vr,t)$ can be written as \cite{joachain_atoms_2011,byron_jr_laser-assisted_1987}:
\begin{widetext}
\begin{equation}
\psi_{nlm}(\vr,t)=\e^{-i E_n t}\left[\phi_{nlm}(\vr)-\frac{A_L}{2}\sum_{n'l'm'}\left(\frac{\e^{i\omega_L t}}{\omega_{n'n}+\omega_L}+\frac{\e^{-i\omega_L t}}{\omega_{n'n}-\omega_L}\right)\langle\phi_{n'l'm'}\vert\veps_L\cdot\hat{\vp}\vert\phi_{nlm}\rangle\phi_{n'l'm'}(\vr)\right],
\label{eq:dressed}
\end{equation}
\end{widetext}
where in the second term we restricted the summation to the bound state solutions $\phi_{nlm}$ of the Schr\"odinger equation for hydrogen like ions and $\omega_{n'n}=E_n-E_{n'}$ is the energy difference between the levels with principal quantum number $n$ and $n'$. Additionally we assume that the laser is not in resonance with any atomic transition.

Even though Eq. \eqref{eq:dressed} holds for any bound state we restrict ourselves to the $1s$ ground state, which is the dominant decay channel. Moreover if this capture into the K-shell is assisted by an optical laser with frequency $\omega_L\ll\SI{10.2}{eV}\leq\omega_{n'1}$ the so called \emph{soft photon approximation} is applicable \cite{reiss_effect_1980,francken_electronatomic-hydrogen_1987} and we can neglect $\omega_L$ in the denominators in Eq. \eqref{eq:dressed}:
\begin{widetext}
\begin{equation}
\psi_{1s}(\vr,t)\equiv\psi_{100}(\vr,t)=\e^{-i E_n t}\left(\phi_{100}(\vr)-A_L \cos (\omega_L t) \sum_{n'l'm'}\frac{1}{\omega_{n'1}}\langle\phi_{n'l'm'}\vert\veps_L\cdot\hat{\vp}\vert\phi_{100}\rangle\phi_{n'l'm'}(\vr)\right).
\label{eq:dressed interm}
\end{equation}
\end{widetext}
To further simplify this expression we apply the Heisenberg equation of motion for the unperturbed atomic Hamiltonian $\hat{H}_0=\hat{\vp}^2/2-Z/r$ and rewrite the matrix element
\begin{equation}
\langle\phi_{n'l'm'}\vert\veps_L\cdot\hat{\vp}\vert\phi_{nlm}\rangle=i\omega_{n'n}\langle\phi_{n'l'm'}\vert\veps_L\cdot\vr\vert\phi_{nlm}\rangle
\end{equation}
from velocity to length form. This allows us to perform the summation in Eq. \eqref{eq:dressed interm} over the states $\phi_{n'l'm'}$ explicitly and to finally obtain:
\begin{equation}
\psi_{1s}(\vr,t)=\e^{-iE_1t}\phi_{100}(\vr)\left[1-iA_L\veps_L\cdot\vr\cos(\omega_Lt)\right].
\label{eq:dressed soft photon}
\end{equation}
The first term on the right hand side of Eq. \eqref{eq:dressed soft photon} is the unperturbed atomic wave function while the second term (\emph{dressing} term) is the perturbative correction due to the external laser. This correction is proportional to $A_Lr$ and, hence, the approximation is valid as long as $A_Lr\ll1$. In our calculations for hydrogen we consider a characteristic length $r=a_0=\SI{1}{a.u.}$.
\subsection{Evaluation of the S-matrix element}
\label{ssec:S-matrix}
With the wave functions before \eqref{eq:continuum} and after \eqref{eq:dressed soft photon} the capture in hand we can calculate the S-matrix of LARR. By inserting these wave functions into Eq. \eqref{eq:sgeneral} we obtain for the recombination into the ground state
\begin{widetext}
\begin{equation}
\begin{aligned}
S_{1s}&=-i\int_{-\infty}^{\infty}\dd t \langle\psi_{1s}(\vr,t)\vert\e^{i\omega_k t}(\veps^{\lambda}_k\cdot\nabla)\vert\phi_{\vp}(\vr,t)\rangle\\
&=-i\int_{-\infty}^{\infty}\dd t\int_{\mathbb{R}^3}\dd^3\vr\e^{i(E_1-E_i+\omega_k)t}\left(\e^{i \kappa \sin(\omega_L t)}+i A_L\veps_L\cdot\vr\cos (\omega_L t)\e^{i \kappa \sin(\omega_L t)}\right)\phi^*_{100}(\vr)(\veps^{\lambda}_k\cdot\nabla)\phi_{\vp}(\vr),
\end{aligned}
\label{eq:stimesplit}
\end{equation}
\end{widetext}
where we defined $\kappa=A_L\veps_L\cdot\vp/\omega_L$. Moreover we applied the dipole approximation for the interaction of the emitted recombination photon and the electron. In this approximation is $\vk\cdot\vr\ll 1$ which is well justified for all cases discussed in this paper.

The time integration in the second line of Eq. \eqref{eq:stimesplit} can be performed analytically if we make use of the Jacobi-Anger expansion and its first derivative with respect to $t$:
\begin{subequations}
\begin{align}
\e^{i\kappa\sin(\omega_L t)}&=\sum_{N=-\infty}^{\infty}\mathrm{J}_N(\kappa)\e^{iN\omega_L t},\\
\cos\omega_L t\e^{i\kappa\sin(\omega_L t)}&=\frac{1}{\kappa}\sum_{N=-\infty}^{\infty}N\,\mathrm{J}_N(\kappa)\e^{iN\omega_L t},
\end{align}
\end{subequations}
where $\mathrm{J}_N(\kappa)$ is the Bessel function of the first kind. The time integrated S-matrix for the recombination into the ground state then can be written as:
\begin{equation}
\begin{aligned}
S_{1s}=-2\pi i\sum_{N=-\infty}^{\infty}&\delta(E_1-E_i+\omega_k-N\omega_L)\\
\times&M^{(N)}(A_L,\veps_L,\vp,Z),
\end{aligned}
\label{eq:smatrix}
\end{equation}
where the delta function provides for energy conservation and $E_1$ is the binding energy of the ground state of the hydrogen-like ion. In this notation the \emph{photon number} $N$ can be interpreted as the number of laser photons that is absorbed ($N>0$) or emitted ($N<0$) by the electron. Therefore the S-matrix \eqref{eq:smatrix} is an infinite sum of partial matrix elements $M^{(N)}(A_L,\veps_L,\vp,Z)$ each corresponding to a particular number $N$ of exchanged laser photons. These partial matrix elements can be written
\begin{widetext}
\begin{equation}
M^{(N)}(A_L,\veps_L,\vp,Z)=\,\mathrm{J}_N(\kappa)\left(M_{RR}(\vp,Z)+\frac{N\omega_L}{\veps_L\cdot\vp}M_{dr}(\vp,Z)\right)
\label{eq:partial me}
\end{equation}
as the sum of two \emph{constituent matrix elements} $M_{RR}(\vp,Z)$ and $M_{dr}(\vp,Z)$
\begin{subequations}
\begin{align}
M_{RR}(\vp,Z)&=B\veps^{\lambda}_k\cdot\vec{e}_p\frac{p-iZ}{(p^2+Z^2)^2}\left(\frac{i \frac{Z}{p}-1}{i \frac{Z}{p}+1}\right)^{iZ/p}\label{eq:M1},\\
M_{dr}(\vp,Z)&=B\left(1-i\frac{Z}{p}\right)\frac{(Z-ip)^{-2iZ/p}}{(p^2+Z^2)^{2-iZ/p}}\left(\veps_L\cdot\veps^{\lambda}_k-2(\veps_L\cdot\vec{e}_p)(\veps^{\lambda}_k\cdot\vec{e}_p)\frac{2p^2-ipZ}{Z^2+p^2}\right),
\end{align}
\label{eq:constituent}
\end{subequations}
where $B=2\pi i\sqrt{2 Z^5}\exp (\pi Z/2p)\Gamma (1-iZ/p)$ and $\vec{e}_p=\vp/p$ is the unit vector that points the initial asymptotic electron momentum.
\end{widetext}

Eq. \eqref{eq:M1} represents the standard matrix element of laser-free radiative recombination, where $A_L=0$. In this case is $\kappa=0$ and we get $S_{1s}=-2\pi i \delta(E_1-E_i+\omega_k) M_{RR}(\vp,Z)$. The constituent matrix element $M_{dr}(\vp,Z)$ has its origin  in the laser dressing contribution to the bound state wave function \eqref{eq:dressed soft photon}. By setting $M_{dr}(\vp,Z)=0$, therefore, we can readily investigate results for bound states uninfluenced by the laser field. This splitting of the partial matrix element $M^{(N)}$ into two parts is consistent with the findings of Shchedrin and Volberg \cite{shchedrin_analytical_2011}.
\subsection{Differential and total LARR cross section}
\label{ssec:cross section}
Up to now we discussed expressions \eqref{eq:partial me} and \eqref{eq:constituent} for the partial matrix element $M^{(N)}$ of laser assisted radiative recombination into the ground state of a hydrogen-like system. We can use these expressions to calculate the total and angle-differential LARR cross section. For example the angle-differential \emph{partial cross section} for the recombination of an electron accompanied by an emission or absorption of $N$ laser photons is given by
\begin{equation}
\frac{\dd\sigma_{1s}^{(N)}}{\dd\Omega_k}=\frac{4\pi^2}{c^3 p}\omega_k(N)\sum_{\lambda}\left| M^{(N)}\right|^2.
\label{eq:differential partial}
\end{equation}
Here $c$ is the speed of light in vacuum and $\Omega_k$ the solid angle into which the recombination photon is emitted. Moreover we assume that the polarization of the recombination photon remains unobserved and hence sum over the helicity $\lambda$.

The frequency $\omega_k$ of the recombination photon in Eq. \eqref{eq:differential partial} is given by the conservation law
\begin{equation}
\omega_k(N)=E_i-E_1+N\omega_L.
\label{eq:recombination photon energy}
\end{equation}
Since $\omega_k$ is always positive, the photon number has to yield the inequality $N\leq N_{min}=(E_i-E_1)/\omega_L$.

From Eq. \eqref{eq:differential partial} we can obtain the total cross section by integrating over the solid angle $\Omega_k$ and summing over the photon number $N$ from $\lceil N_{min}\rceil$ to infinity:
\begin{equation}
\sigma_{1s}=\frac{4\pi^2}{c^3 p}\sum_{N=N_{min}}^{\infty}\sum_{\lambda}\int\dd\Omega_k\omega_k(N)\left| M^{(N)}\right|^2.
\label{eq:total}
\end{equation}
This cross section characterizes the probability to produce a hydrogen ion by LARR. However the sum in Eq. \eqref{eq:total} is infinite not all terms contribute equally to the total cross section. If we perform a stationary phase analysis of Eq. \eqref{eq:stimesplit} (following Refs. \cite{seipt_laser-assisted_2014,kuchiev_multiphoton_2000}) we find that the partial matrix element $M^{(N)}$ is nearly zero if $|N|$ exceeds the \emph{cut off photon number}
\begin{equation}
N_{cut}=A_L p/\omega_L.
\label{eq:ncut}
\end{equation}
Accordingly if $N_{cut}<|N_{min}|$ we can take $N_{min}\rightarrow-\infty$ without changing the result significantly.
\section{Results and discussion}
\label{sec:results and discussion}
\subsection{Differential partial cross section}
\label{ssec:results differential}
\begin{figure}[tb]
\centering
\includegraphics[width=\linewidth]{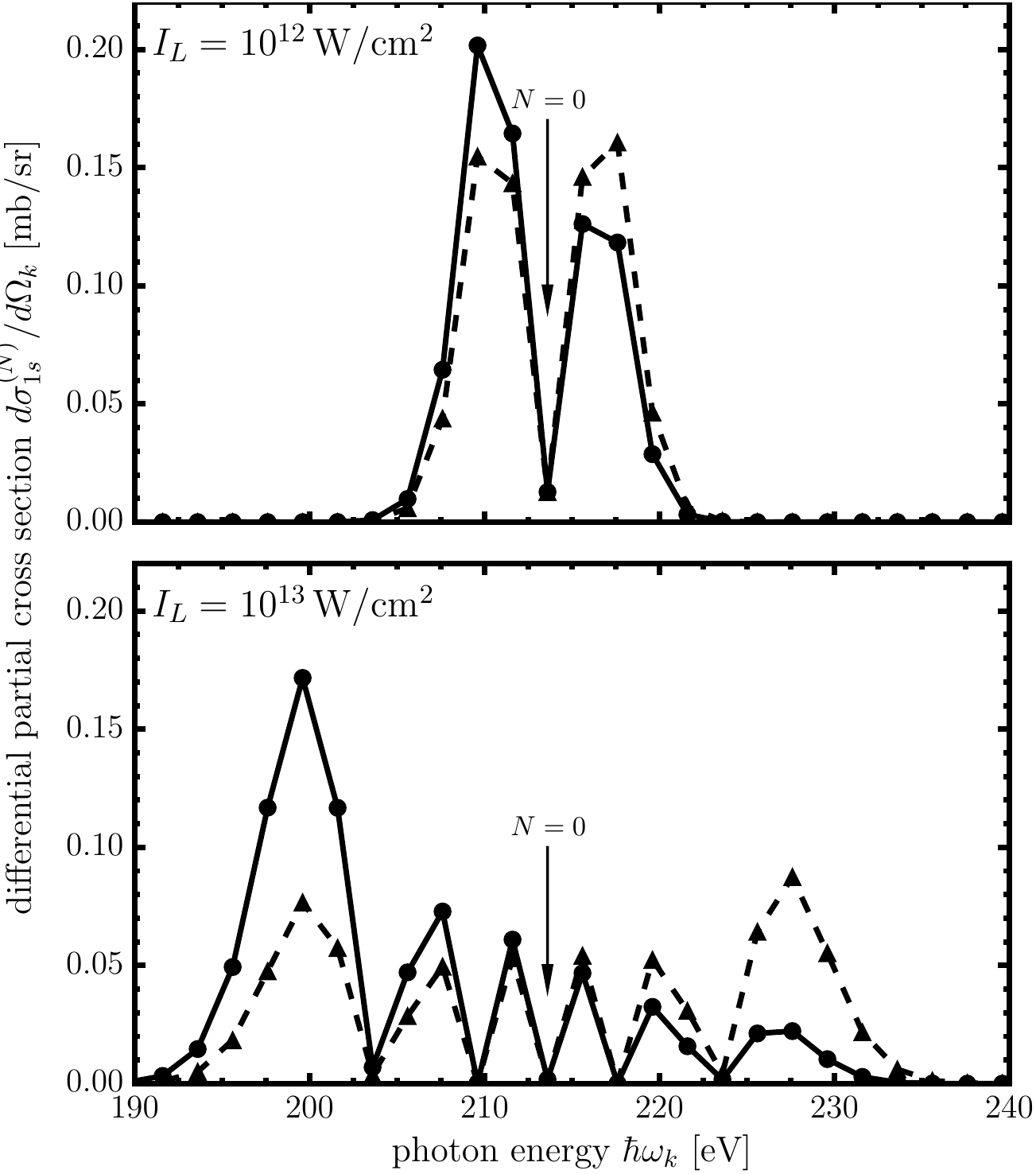}
\caption{Angle-differential partial cross section \eqref{eq:differential partial} for the laser assisted recombination of a $E_i=\SI{200}{eV}$ electron and a hydrogen nucleus ($Z=1$) as a function of the recombination photon energy $\omega_k(N)$. The calculations have been performed for a photon emission angle $\theta_k=175^\circ$ and a laser polarization angle $\chi=135^\circ$. The laser parameters are $\hbar\omega_L=\SI{2}{eV}$, $I_L=\SI{e12}{W/cm^2}$ (upper panel) and $I_L=\SI{e13}{W/cm^2}$ (lower panel). The dashed lines and triangles refer to the results where the dressing of the target bound states is neglected. The spectrum is discrete and the lines are only shown to guide the eye.\label{fig:angular_spectrum}}
\end{figure}
\begin{figure*}[tb]
\centering
\includegraphics[width=\linewidth]{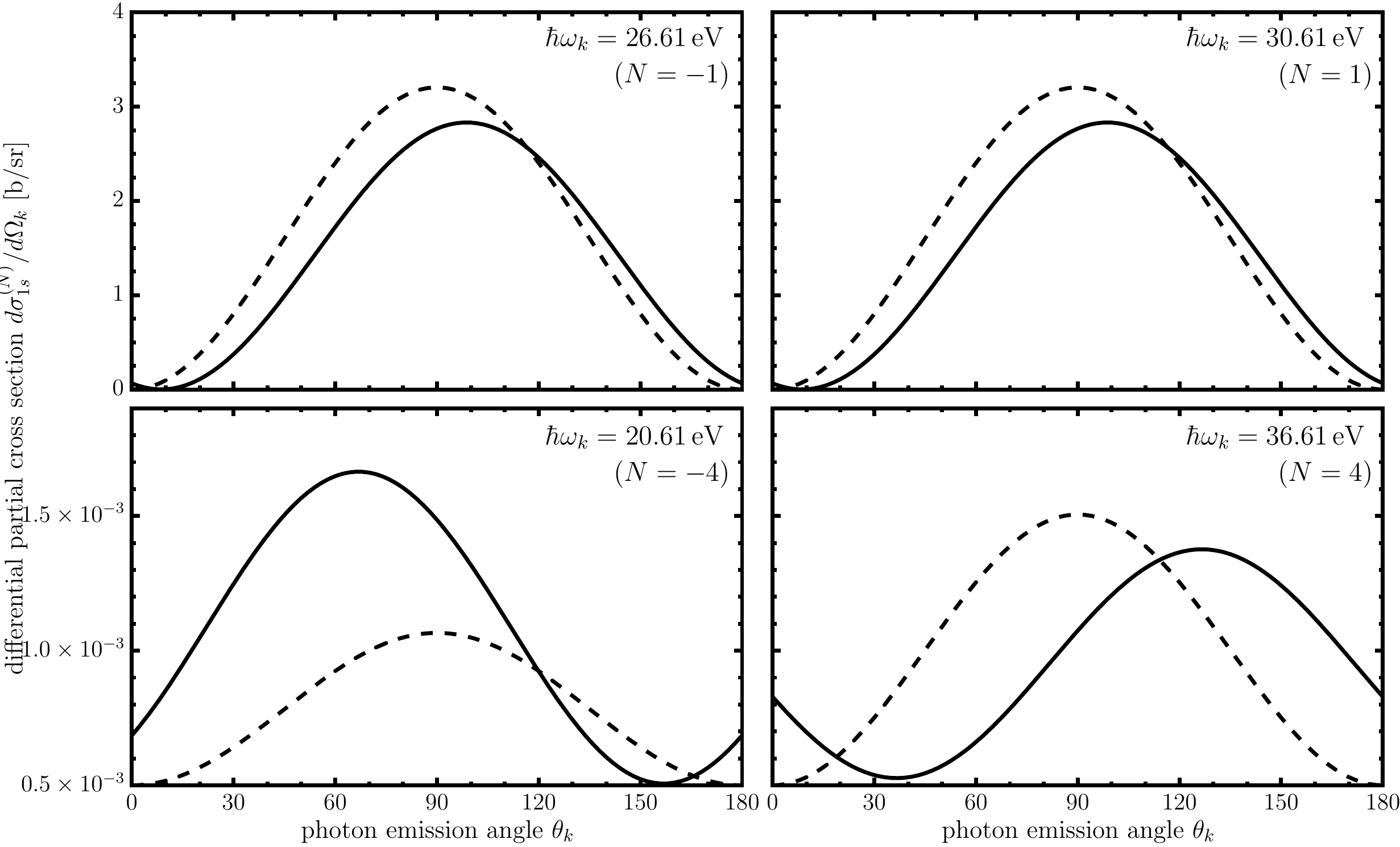}
\caption{Angular differential partial cross section for the laser assisted recombination of a $E_i=\SI{15}{eV}$ electron into the ground state of hydrogen (Z=1) as a function of the photon emission angle $\theta_k$. The assisting laser has an intensity of $I_L=\SI{e13}{W/cm^2}$ and the frequency $\hbar\omega_L=\SI{2}{eV}$. Its polarization angle is $\chi=65^\circ$. Results are shown for four different recombination photon energies; in the upper row $\hbar\omega_k(N=-1)=\SI{26.61}{eV}$ (left panel) and $\hbar\omega_k(N=1)=\SI{30.61}{eV}$ (right panel) while in the lower row $\hbar\omega_k(N=-4)=\SI{20.61}{eV}$ (left panel) and $\hbar\omega_k(N=4)=\SI{36.61}{eV}$ (right panel). In all panels we show results obtained including (solid lines) and omitting (dashed lines) the bound state dressing in comparison.\label{fig:angular_differential}}
\end{figure*}
In the past LARR has been discussed mainly within the strong field approximation (SFA) \cite{becker_modeling_1994, jaron_stimulated_2000}. The laser influence on the bound state is neglected in this approximation. Anyhow we expect from previous attempts \cite{shchedrin_analytical_2011} that the emission pattern and energy distribution of the recombination photons is affected by the bound state dressing. Therefore we use our theory to investigate these effects and perform calculations for the angle-differential partial cross section.

In Fig. \ref{fig:angular_spectrum} we show the differential partial cross section of laser assisted recombination into the $1s$ state of hydrogen. The calculations have been carried out for a fixed photon emission angle $\theta_k=175^\circ$ and an external laser ($\hbar\omega_L=\SI{2}{eV}$) with two different laser intensities $I_L=\SI{e12}{W/cm^2}$ (upper panel) and $I_L=\SI{e13}{W/cm^2}$ (lower panel). Results are shown with (circles) and without (triangles) incorporation of the laser dressing of the bound state. Of course,  the LARR cross section is only defined at the discrete energies $\omega_k(N)$ shown in Eq. \eqref{eq:recombination photon energy} and the solid and dashed lines are only shown to guide the eye.

It is seen in Fig. \ref{fig:angular_spectrum} that the spectrum is not just a single line, as it would be for the laser-free radiative recombination, but a distribution of photon energies. The width of this distribution increases from about $\SI{18}{eV}$ for $I_L=\SI{e12}{W/cm^2}$ to $\SI{46}{eV}$ for $I_L=\SI{e13}{W/cm^2}$. This intensity-dependent broadening of the spectrum can be understood from the scaling of cut off photon number $N_{cut}$ that becomes larger as the intensity increases (cf. Eq. \eqref{eq:ncut}).

As seen from Fig. \ref{fig:angular_spectrum} the "dressed" and "undressed" results behave quite differently as a function of $\omega_k(N)$. The results obtained without bound state dressing are almost symmetric around the field free photon energy where $N=0$. A slight increase of $\dd\sigma^{(N)}_{1s}/\dd\Omega_k$ towards higher photon energies is caused by the prefactor $\omega_k(N)$ in Eq. \eqref{eq:differential partial} that scales linearly but weakly with $N$. If, in contrast, the bound state dressing is included into the computations, the photon distribution becomes strongly asymmetric. This can be explained by the fact that the contribution of $M_{dr}(\vp,Z)$ increases with $N$ as seen from Eq. \eqref{eq:partial me}. For higher intensities sidebands with larger photon numbers are visible and therefore the asymmetries become more pronounced.

Fig. \ref{fig:angular_spectrum} displays the spectral distribution \eqref{eq:differential partial} of the recombination photons for a fixed photon emission angle $\theta_k$. In order to investigate the $\theta_k$-dependence of $\dd\sigma^{(N)}_{1s}/d\Omega_k$ we present in Fig. \ref{fig:angular_differential} the angular distribution of the recombination photons emitted during the recombination of $\SI{15}{eV}$ electrons into the $1s$ state of hydrogen. The calculations were performed for a laser with intensity $I_L=\SI{e13}{W/cm^2}$, frequency $\hbar\omega_L=\SI{2}{eV}$ and different energies of the recombination photons: $\hbar\omega_k(N=-1)=\SI{26.61}{eV}$ (top left panel), $\hbar\omega_k(N=1)=\SI{30.61}{eV}$ (top right panel), $\hbar\omega_k(N=-4)=\SI{20.61}{eV}$ (bottom left panel) and $\hbar\omega_k(N=4)=\SI{36.61}{eV}$ (bottom right panel). Again we compare results obtained including (solid lines) and neglecting (dashed lines) the bound state dressing. As seen from Fig. \ref{fig:angular_differential}, the emission pattern of the recombination photons is symmetric around $90^\circ$ if the bound state wave function is assumed to be uninfluenced by the laser. This behaviour is easily understood by setting $M_{dr}(\vp,Z)=0$ in Eq. \eqref{eq:partial me}. The LARR cross section \eqref{eq:differential partial} can be written then as $\dd\sigma_{1s}^{(N)}\sim\sum_\lambda|\veps_k^\lambda\vec{e}_p|^2=1-|\vec{e}_p\vec{e}_k|^2=\sin^2\theta_k$ (see Ref.\cite{landau_quantum_1991}). This angular distribution is independent on the photon number $N$ and, therefore, the same for all recombination photon energies (cf. Fig. \ref{fig:angular_differential}). However a different angular behaviour occurs if the bound state dressing is taken into account. The constituent matrix element $M_{dr}(\vp,Z)$ contains additional angular dependent terms that lead to an asymmetric shift of the angular distribution. This shift becomes larger for higher photon numbers $N$ which enter Eq. \eqref{eq:partial me} as a prefactor of $M_{dr}(\vp,Z)$. However as seen from Fig. \ref{fig:angular_spectrum} and Eq. \eqref{eq:ncut} large N contributions become only visible if $I_L$ or $E_i$ is sufficiently high.
\subsection{Total cross section of LARR}
\label{ssec:results total}
\begin{figure}[tb]
\centering
\includegraphics[width=\linewidth]{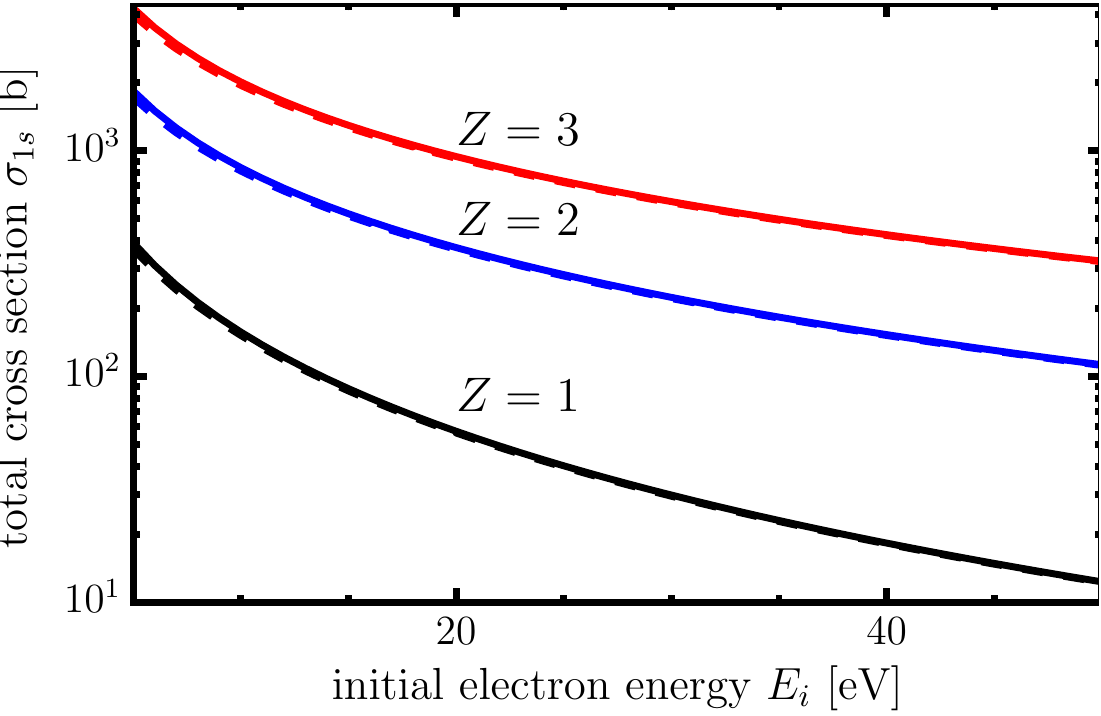}
\caption{Total cross section of laser assisted radiative recombination into the ground state of a hydrogen-like ion as a function of the initial electron energy $E_i$ for three different nuclear charges (Z=1: black, Z=2: blue, Z=3: red). The calculations have been performed for a laser with the intensity $I_L=\SI{e13}{W.cm^{-2}}$, laser photon energy $\hbar\omega=\SI{2}{eV}$ and a polarization angle $\chi=90^\circ$. We show results calculated with laser dressed bound state wave functions (solid lines) and results where the bound state dressing is neglected (dashed lines).\label{fig:vary_Ei}}
\end{figure}
\begin{figure}[tb]
\centering
\includegraphics[width=\linewidth]{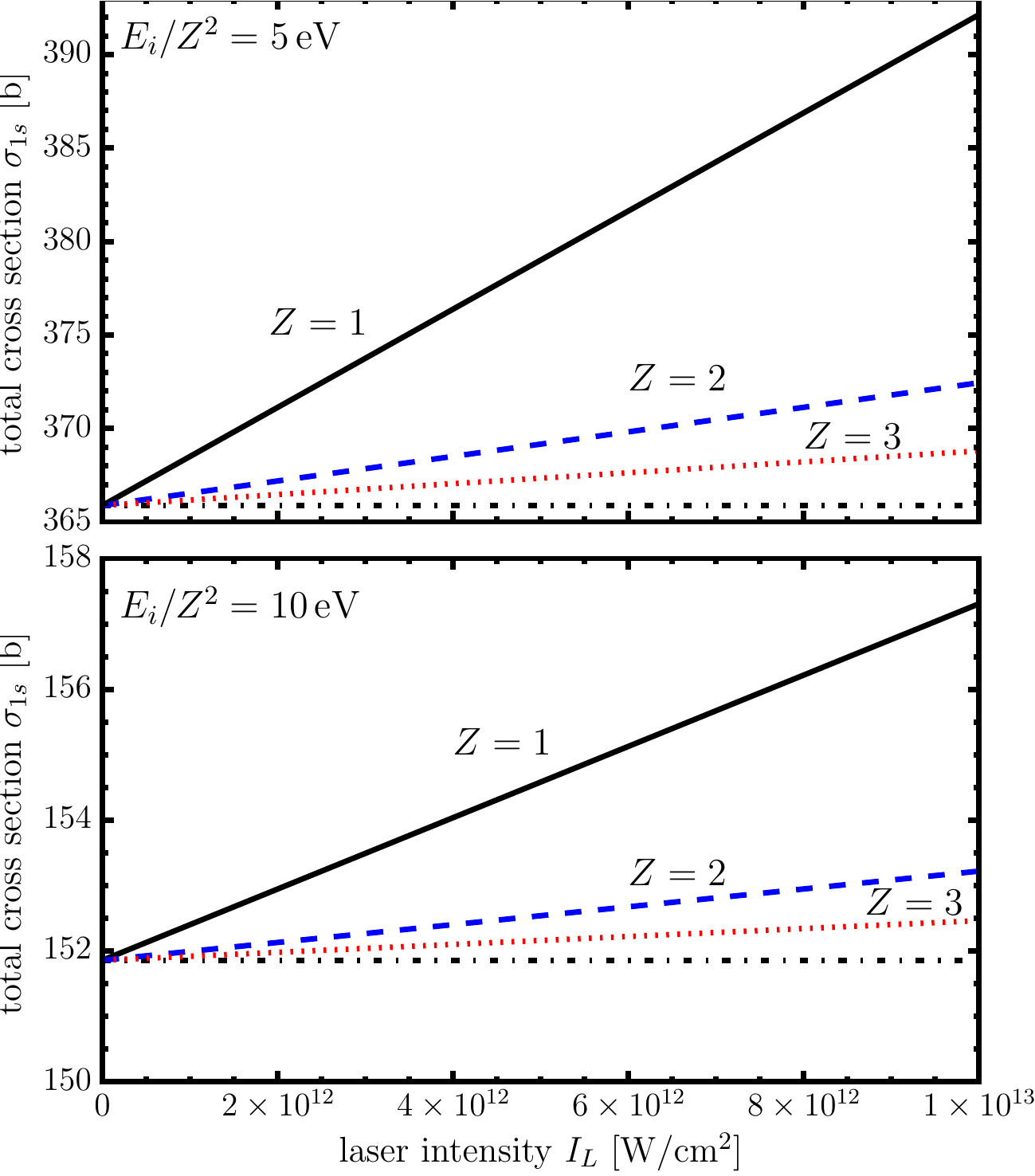}
\caption{Total cross section for laser assisted radiative recombination into the ground state of a hydrogen-like ion as a function of the laser intensity $I_L$ for $\omega_L=\SI{2}{eV}$ and three different nuclear charges (Z=1: black, solid; Z=2: blue, dashed; Z=3: red, dotted).The initial electron energy is normalized with respect to the nuclear charge such that $E_i/Z^2=\SI{5}{eV}$ (upper panel) and $E_i/Z^2=\SI{10}{eV}$ (lower panel). The polarization angle is $\chi=90^\circ$. For comparison the laser free case, respectively the result obtained with undressed bound state wave functions is shown (black dot-dashed line).\label{fig:vary_EL}}
\end{figure}
In the previous section we have shown that the differential partial cross section \eqref{eq:differential partial} of laser assisted radiative recombination (LARR) may be influenced remarkably by the laser dressing of the residual bound atomic state. This effect becomes more pronounced for higher photon numbers $N$. As we have shown in Sec. \ref{ssec:cross section} however only the channels with $|N|\leq N_{cut}$ contribute to the cross section. Since $N_{cut}$ depends on the energy of the incident electron $E_i$ and the laser intensity $I_L$ we expect that also the total cross section as a function of $E_i$ and $I_L$ is influenced by the bound state dressing. In order to study this we display in Fig. \ref{fig:vary_Ei} the total cross section \eqref{eq:total} of LARR as a function of the incident electron energy for three different nuclear charges $Z$. For our calculations we set the laser intensity $I_L=\SI{e13}{W/cm^2}$ and the frequency $\hbar\omega_L=\SI{2}{eV}$. Again we compare results obtained with the laser dressing being neglected (dashed lines) and taken into account (solid lines). From this comparison we see that in the latter case the LARR cross section is enhanced. This difference between the two cases (with and without bound state dressing) vanishes for increasing kinetic energies of the incident electron. For example for $Z=1$ in the low energy regime at $E_i=\SI{5}{eV}$, $\sigma_{1s}$ is increased by a factor of $1.07$. For $E_i=\SI{30}{eV}$ this factor is only $1.01$.

Up to now we discussed the total LARR cross section $\sigma_{1s}$ as a function of the initial electron energy $E_i$ for different nuclear charges $Z$ and $I_L=\SI{e13}{W/cm^2}$. Beyond that we expect that $\sigma_{1s}$ can also depend on the laser intensity. Figure \ref{fig:vary_EL} shows the total LARR cross section \eqref{eq:total} as a function of the laser intensity $I_L$, again for three different nuclear charges $Z$. Moreover we present our results for two sets of initial electron energies $E_i=\SI{5}{eV}\cdot Z^2$ and $E_i=\SI{10}{eV}\cdot Z^2$. We apply this scaling of $E_i$ to better illustrate how the dressing effects depend on $Z$. Because the laser free cross section is defined solely by the Sommerfeld parameter $\nu=\sqrt{Z^2/(2E_i)}$, the $Z$ dependence of the total cross section at a fixed $E_i/Z^2$ comes from the bound state dressing only. Similarly to before we see in Fig. \ref{fig:vary_EL} that the effect of bound state dressing decreases with increasing energy $E_i$. Moreover the effect of bound state dressing is decreased for higher nuclear charges $Z$. This is due to the stronger binding of the electrons in the ground state of the hydrogen like ion.

It can be seen in Fig. \ref{fig:vary_EL} that the bound state dressing causes a qualitative different behaviour of the total LARR cross sections. The results obtained without bound state dressing are independent on the laser intensity while the full calculations increase linearly with $I_L$. To explain this behaviour we insert the partial matrix element \eqref{eq:partial me} into Eq. \eqref{eq:total} and perform the summation over $N$. Since for the discussed scenario $N_{cut}<|N_{min}|$ is satisfied the sum can be extended to negative infinity. This enables us to use
\begin{subequations}
\begin{align}
\sum_{N=-\infty}^{\infty}\mathrm{J}^2_N(\kappa)&=1,\\
\sum_{N=-\infty}^{\infty}N^2\mathrm{J}^2_N(\kappa)&=\frac{\kappa^2}{2},\\
\sum_{N=-\infty}^{\infty}N^{2j+1}\mathrm{J}^2_N(\kappa)&=0\text{ for }j\in \mathbb{N}_0,
\end{align}
\end{subequations} 
and to obtain the total cross section in the form
\begin{equation}
\begin{aligned}
\sigma_{1s}&=\frac{4\pi^2}{c^3p}\sum_{\lambda}\int\dd\Omega_k\biggl[(E_i-E_1)|M_{RR}|^2\\
&+\frac{E_L^2}{2\omega_L^2}\big( (E_i-E_1)|M_{dr}|^2+(\veps_L\cdot\vp)\mathrm{Re}(M_{RR}^*M_{dr})\big)\biggr].
\end{aligned}
\label{eq:scaling}
\end{equation}
From this expression we see that the total cross section scales quadratically with the laser electric field amplitude $E_L$ and therefore linearly with the laser intensity $I_L$. If in contrast the bound state dressing is neglected ($M_{dr}=0$), the total cross section $\sigma_{1s}$ does not depend on $I_L$. Moreover in this case Eq. \eqref{eq:scaling} reduces to the standard "laser-free" RR cross section. This implies that, within the present model, the assisting laser affects the total cross section solely via the bound state dressing.
\section{Summary}
\label{sec:summary}
In this paper we presented a theoretical study on the recombination of a free electron into the ground state of a hydrogen-like ion in the field of an external laser. We put special emphasis on the effects which arise from the laser dressing of the bound ionic state. In order to investigate how this dressing affects the total and angle-differential LARR cross section we applied the S-matrix theory and the dipole approximation for the coupling of the electron to electromagnetic fields. Within this theory the incident electron in the combined field of the laser and the nucleus is described by a separable Coulomb-Volkov wavefunction. For the influence of the laser field on the final bound state we accounted using time dependent perturbation theory up to the first order.

Based on the developed approach we performed calculations for the differential partial cross section $\dd\sigma^{(N)}_{1s}/\dd\Omega_k$ as a function of the emission angle $\theta_k$ and the energy $\omega_k(N)$ of the recombination photon. We found that $\dd\sigma^{(N)}_{1s}/\dd\Omega_k$ is symmetric around $\theta_k=90^\circ$ if the bound state dressing is neglected while the full calculations show an asymmetric shift of the angular distribution. Comparing the energy distribution of the emitted recombination photons obtained either with or without bound state dressing we see remarkable differences especially for higher photon numbers $N$. 

Moreover results have been obtained for the total cross section as a function of the incident electron energy $E_i$ and the laser intensity $I_L$. We found that, within the validity of our theory, the total cross section of LARR may be enhanced if the bound state dressing is taken into account. It was shown analytically that this enhancement is linear in $I_L$. However if the bound state dressing is neglected the total LARR cross section is constant with respect to $I_L$ and coincides with the total cross section of laser free radiative recombination.

Laser assisted atomic processes gained rising interest for the recent years especially using intense laser fields. While it has been common to neglect dressing effects on bound atomic states this study stresses their importance for higher intensities. 
\begin{acknowledgments}
RAM acknowledges support of the Studienstiftung des Deutschen Volkes.
\end{acknowledgments}
\appendix
\section{Derivation of the constituent matrix elements}
\label{app:constituent}
To derive closed expressions for the constituent matrix elements we insert the wave functions $\psi_{\vp}(\vr,t)$ \eqref{eq:continuum} and $\psi_{1s}(\vr,t)$ \eqref{eq:dressed soft photon} into the general equation for the S-matrix \eqref{eq:stimesplit}. After time integration we find:
\begin{subequations}
\begin{align}
M_{RR}(\vp,Z)&=\int_{\mathbb{R}^3}\dd^3\vr \phi_{100}(\vr)(\veps_k\cdot\nabla)\psi_{\vp}(\vr),\\
M_{dr}(\vp,Z)&=\int_{\mathbb{R}^3}\dd^3\vr\veps_L\cdot\vr \phi_{100}(\vr)(\veps_k\cdot\nabla)\psi_{\vp}(\vr).
\end{align}
\end{subequations}
To perform the spatial integration analytically we multiply the integrands by $1=e^{i\vk\cdot\vr}|_{\vk=0}$ and obtain:
\begin{subequations}
\begin{align}
M_{RR}(\vp,Z)&=2^6\pi^3B(\veps_k\cdot\nabla_k)\left.\mathcal{I}(Z,\vp,\vk)\right|_{\vk=0},\\
M_{dr}(\vp,Z)&=-i2^6\pi^3B(\veps_L\cdot\nabla_k)(\veps_k\cdot\nabla_k)\left.\mathcal{I}(Z,\vp,\vk)\right|_{\vk=0},
\end{align}
\label{eq:constituent_simplified}
\end{subequations}
where $B$ has been introduced in Eq. \eqref{eq:constituent} and $\nabla_k$ is the gradient with respect to the components of $\vk$. The remaining integral over the spatial coordinates is now reduced to a standard atomic integral $\mathcal{I}(Z,\vp,\vk)$ that can be found, e.g., in Ref.\cite{landau_quantum_1991} yielding
\begin{widetext}
\begin{equation}
\mathcal{I}(Z,\vp,\vk)=\int_{\mathbb{R}^3}\frac{\dd^3\vr}{r}\e^{i(\vp-\vk)\cdot\vr-Zr}\,{}_1\mathrm{F}_1(iZ/p,1,i(pr-\vp\cdot\vr))=4\pi\frac{(k^2+(Z-ip)^2)^{-iZ/p}}{((\vp-\vk)^2+Z^2)^{1-iZ/p}}.
\end{equation}
\end{widetext}
Using this relation and performing the $\vk$-derivatives in \eqref{eq:constituent_simplified} we obtain the results shown in Eq. \eqref{eq:constituent}:
\begin{widetext}
\begin{subequations}
\begin{align}
M_{RR}(\vp,Z)&=B\veps^{\lambda}_k\cdot\vec{e}_p\frac{p-iZ}{(p^2+Z^2)^2}\left(\frac{i \frac{Z}{p}-1}{i \frac{Z}{p}+1}\right)^{iZ/p},\\
M_{dr}(\vp,Z)&=B\left(1-i\frac{Z}{p}\right)\frac{(Z-ip)^{-2iZ/p}}{(p^2+Z^2)^{2-iZ/p}}\left(\veps_L\cdot\veps^{\lambda}_k-2(\veps_L\cdot\vec{e}_p)(\veps^{\lambda}_k\cdot\vec{e}_p)\frac{2p^2-ipZ}{Z^2+p^2}\right).
\end{align}
\end{subequations}
\end{widetext}
\bibliography{quellen.bib}
\end{document}